# Integrative Data Semantics through a Model-enabled Data Stewardship

Philipp Wegner[1,†], Sebastian Schaaf[1,†], Mischa Uebachs[2], Daniel Domingo-Fernández[1,3], Yasamin Salimi[1,4], Stephan Gebel[1], Astghik Sargsyan[1,4], Colin Birkenbihl[1,4], Stephan Springstubbe[1], Thomas Klockgether[2,5], Juliane Fluck[1,7], Martin Hofmann-Apitius[1,4], and Alpha Tom Kodamullil[1,8,*]

1) Department of Bioinformatics, Fraunhofer Institute for Algorithms and Scientific Computing (SCAI), Sankt Augustin 53754, Germany

2) Department of Neurology, University Hospital Bonn (UKB), Bonn 53127, Germany

3) Enveda Biosciences, Boulder, CO, 80301, USA

4) Bonn-Aachen International Center for IT (B-IT), Rheinische Friedrich-Wilhelms-Universität Bonn, Bonn 53115, Germany

5) German Center for Neurodegenerative Diseases (DZNE), Bonn 53127, Germany

6) Institute of Geodesy and Geoinformation (IGG), Rheinische Friedrich-Wilhelm-Universität Bonn, Bonn 53115, Germany

7) Department Knowledge Management, Information Centre for Life Sciences (ZBMED), 53115 Bonn

8) Causality Biomodels, Kinfra Hi-Tech Park, Kalamassery, Cochin, Kerala 683503, India

*To whom correspondence should be addressed.

† These authors contributed equally to this work.

## Abstract

**Motivation:**

The importance of clinical data in understanding the pathophysiology of complex disorders has prompted the launch of multiple initiatives designed to generate patient-level data from various modalities. While these studies can reveal important findings relevant to the disease, each study captures different yet complementary aspects and modalities which, when combined, generate a more comprehensive picture of disease aetiology. However, achieving this requires a global integration of data across studies, which proves to be challenging given the lack of interoperability of cohort datasets.

**Results:**

Here, we present the Data Steward Tool (DST), an application that allows for semi-automatic semantic integration of clinical data into ontologies and global data models and data standards. We demonstrate the applicability of the tool in the field of dementia research by establishing a Clinical Data Model (CDM) in this domain. The CDM currently consists of 277 common variables covering demographics (e.g. age and gender), diagnostics, neuropsychological tests, and biomarker measurements. The DST combined with this disease-specific data model shows how interoperability between multiple, heterogeneous dementia datasets can be achieved.

**Availability:**

The DST source code and Docker images are respectively available at https://github.com/SCAI-BIO/data-steward and https://hub.docker.com/r/phwegner/data-steward. Furthermore, the DST is hosted at https://data-steward.bio.sca.fraunhofer.de/data-steward.

**Corresponding author's email:** alpha.tom.kodamullil@scai.fraunhofer.de

# 1. Introduction

Vast amounts of patient data of heterogeneous types are produced on a daily basis. This data can be generated in different locations, including hospitals or primary care centers, in regional multi-center studies, national centers, and international consortia. Typically, while there are multiple studies that collect heterogeneous information from patients having a certain condition, designs of studies and records often do not match. Consequently, the underlying data cannot be directly compared across studies without a semantic harmonization that enables dataset interoperability. This lack of interoperability across datasets prevents one from getting the most out of the data as well as conducting independent validations on external datasets (Birkenbihl *et al*., 2020).

A solution to this problem lies in the establishment of a so-called global clinical data model (**see examples in Supplementary Table 1**) which aims at integrating information across heterogeneous datasets. In doing so, such global clinical data models could be used by cataloging activities, repositories and computational data environments in order to manage data based on schemata and metadata templates, ultimately improving interoperability and exchange across resources. Individual research institutions also stand to benefit as they can organize data according to a common data model in their domain. As such, developing domain-specific common data models can have a catalyst effect that accelerates the sharing and exchange of data that can influence future design of studies. Additionally, existing common data models are not yet capable enough to map all variables from a particular topic, for instance, dementia (**Supplementary Text 1**).

Here, we present the Data Steward Tool (DST) which can be used to automatically standardize clinical datasets, map them to established ontologies, align them with OMOP standards (a widely used data model for health data), and export them to a FHIR-based format. To standardize specific variables from dementia which could not be mapped to existing ontologies or data standards, we developed a Clinical Data Model (CDM) in the context of dementia, built from various datasets - Alzheimer's disease datasets (ADNI (Mueller *et al.,* 2005), and AddNeuroMed (Lovestone *et al*., 2009)), routine clinical datasets (from University Hospitals Bonn and Aachen, Germany) and other datasets covering neurodegenerative diseases (from DZNE (https://www.dzne.de) and EUROSCA (http://www.eurosca.org).

# 2. Materials and Methods

**Implementation**

We implemented a web application called the DST that allows users to read, edit and use the CDM to standardize real-world research data. The tool consists of RESTful APIs served by a Django application on a MongoDB database **(technical details are described in Supplementary Text 2)**; thus, providing a user-friendly environment that makes the underlying data accessible to the scientific community.

To demonstrate the DST, we accessed and investigated major dementia studies (**Supplementary Text 3**) and identified the variables they shared in common. We found that the majority of these variables comprised patient information, such as age, sex or initial diagnosis, biomarker measurements, as well as measurements from neuropsychological tests, such as the Boston Naming Test. This set of variables (**Supplementary Table 2**) was then used as the groundwork to extend the data model as we gained access to new studies by connecting the variables of additional studies to equivalent variables present in the CDM (**Supplementary Text 4**). Finally, we extracted data types and value ranges for each of the variables and assigned every variable to a specific data modality.

## 3. Results and discussion

**Data Steward Tool for Data Standardization**

We developed a web application called the DST that provides an interface to visualize the model, extend it with new variables, add mappings, and read clinical data through a user-friendly interface **(Figure 1A)**. Furthermore, the DST is capable of automatically mapping external variables onto the CDM through fuzzy string matching. Performing full text searches throughout the entire data model grants the possibility of using the system as a searchable variable catalog for dementia research. Additionally, uploading clinical data onto the system allows for data standardization and the storage of harmonized data in a central data repository **(Figure 1B)** The uploaded standardized data can be queried via a RESTful API. Through these APIs, users can access the content of the DST using their own tools (e.g., Jupyter notebook). Uploading and standardizing data is a guided multi-step process (**detailed description in Supplementary Text 5**), where the user begins by loading the data, often a 2D data table, to an interface via drag and drop and is then given the opportunity to manually map variables that were not found in the current data model **(Figure 1C)**. During this process, the user is supported with autosuggestions to automatically add terms from OLS if no suitable variable could be found. Finally, the DST provides a graph-based view (**Figure 1D**) of the model where the user can interactively explore the entirety of the model.

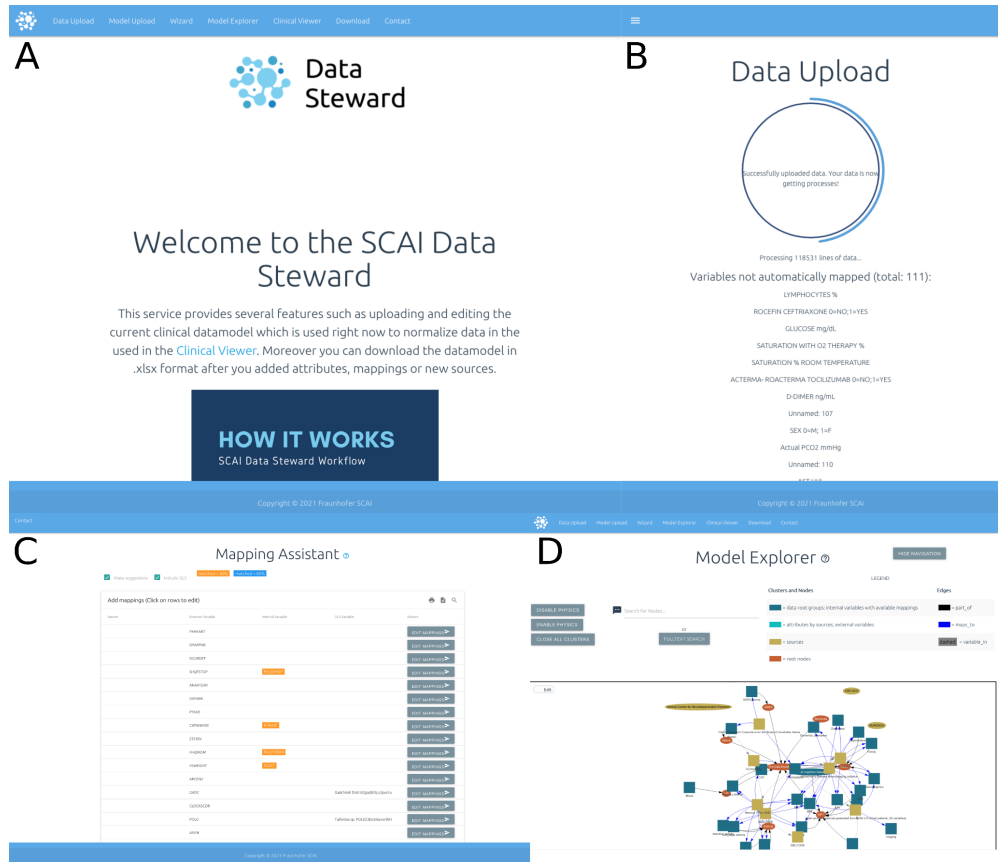

**Figure 1**. Overview of Data Steward Tool features. A) Home page. B) Data upload page. C) Mapping assistant view. D) Data Model explorer view.

To make the DST fully compatible with existing approaches currently used in industry, the tool functions as a FHIR node that is capable of providing patient data in FHIR format. Furthermore, the data model is aligned with OMOP (**Supplementary Text 6**) and can easily be enhanced with existing data models (**Supplementary Text 7**).

## Clinical Data Model for Dementia

The CDM consists of 277 common variables representing cohort demographic information, various clinical assessments, and further dementia-related biomarkers. The CDM was designed to store metadata about the variables themselves, including definitions, data type, and value ranges as well as information about the variable names already mapped onto it. The CDM can constantly and rapidly expand as more studies are mapped to it. By enriching the variables present in the data model with mappings to other resources, the system is capable of standardizing data from different origins. Finally, while the data model is stored in a database, it can be exported to tabular formats (e.g., excel and csv), OWL, and RDF.

## Clinical Viewer

The DST is accompanied with the Clinical Viewer (https://data-steward.bio.scai.fraunhofer.de/), a complementary visualization tool to explore the clinical data stored in the DST using various visualization techniques such as scatter plots, pie charts, and histograms. In order to demonstrate the DST and the Clinical Viewer, we used publicly available patient data to generate 10 virtual patients similar to those of the ADNI dataset (Mueller *et al.,* 2005). These patients were then duplicated until we had datasets of 1,000 and 10,000 patients (**see details in Supplementary Text 8**). The synthetic 1,000 patients ADNI dataset is currently used as an example in the demo instances of the DST.

## 4. Conclusion and Future Work

The Data Steward Tool is a web application designed to bring a convenient and easy-to-use solution for the semantic integration of clinical data. By design, the system's backend functions as a generic solution for data standardization and provides the possibility to map cohort datasets to global data models and ontologies. Moreover, the DST has various applications like the Clinical Viewer or could act as the data pre-processor for machine learning tasks. Any new dataset can be described and qualified in relation to the common data model, making it instantly clear what the strengths and weaknesses of a dataset are. For example, a common clinical data model would implicitly allow for an assessment of the coverage (with respect to variables) of a new study by mapping the new study to the common data model. As further steps, we plan to enrich the DST with more complex query APIs to make the tool usable for future data science tasks in bioinformatics. In order to further improve the data model, we plan to add more variables and mappings from diverse data sources and clinical studies.

## Funding

This work has been supported by the IDSN project (project number: 031L0029 [A-C]).

## Competing Interests

# Supplementary Information

## Supplementary Tables

| Name | Description | Source |
|---|---|---|
| OMOP | Data model for observational health data | https://www.ohdsi.org/data-standardization/ |
| FHIR | **F**ast **H**ealthcare **I**nteroperability **R**esources is a standard for health care data exchange | https://www.hl7.org/fhir/overview.html |
| Ga4GH | GA4GH provides a tool kit for standardized genomic data sharing | https://ga4gh.github.io/tool-registry-service-schemas/DataModel/ |
| I2b2 | i2b2 provides software for structured integration / normalization of health data | https://www.i2b2.org/ |

**Supplementary Table 1. Global data models.** Examples of global clinical data models.

| Variable name | Short description |
|---|---|
| SEX | Gender |
| APOE_STATUS | ApoE Risk group |
| DOB | Date of birth |
| RW_BNTTOT | Raw score of Boston Naming Test |
| RW_CERCP | Constructive Practice - direct (raw score) |
| RW_CERCPR | Constructive Practice - delayed (raw score) |
| RW_CERCPSAV | Constructive Practice - Savings (raw score) |
| RW_CERDISC | Word list - Discriminability (raw score) |
| RW_CERDRLCT | Word list - Retrieval after interference (raw score) |
| RW_CERDRLIT | Word list - Intrusions (Raw Score) |
| RW_CERRL | Word list - sum (raw value) |
| RW_CERRL1CT | Word list - 1st round (raw score) |
| RW_CERRL2CT | Word list - 2nd round (raw score) |
| RW_CERRL3CT | Word list - 3rd round (raw score) |
| RW_CERWLSAV | Word list - Savings (raw value) |
| RW_MMSTOT | Mini Mental State Test (raw score) |
| RW_PFSTOT | Verbal fluency phonetically (raw score) |
| RW_TMTA | TMT Boy Scout Test - Part A (raw score) |
| RW_TMTB | TMT Boy Scout Test - Part B (raw score) |
| RW_TMTBA | TMT Boy Scout Test - B / A (raw score) |
| RW_VFATOT | Verbal fluency cat. (Raw value) |
| LIQ_A40_RESN | CSF Amyloid beta 40 |
| LIQ_A42_A40_RATIO | CSF Ratio 42/40 |
| LIQ_A42_RESN | CSF Amyloid beta 42 |
| LIQ_LEU_RESN | CSF leukocytes |
| LIQ_PTP_RESN | CSF pTau |
| LIQ_TP_C_RESN | CSF Protein (total) |
| LIQ_TTP_RESN | CSF Tau (total) |
| LIQ_A38_RESN | CSF Amyloid beta 38 |
| MEMORY | Memory impairment |

| LANGUAGE | Speech disorder |
|---|---|
| DIAGNOSIS | Diagnosis |
| LIQ_A42_A40_RATIO_CALC | CSF Ratio 42/40 |
| SCACAT | Patient category (SCA known, SCA unknown, control) |
| SCACAT2 | SCA genotype |
| CARRIER | Mutation carrier vs. first-degree relative of mutation carrier |
| ATAXIC | Ataxic yes / no |
| AOO | Age at first gait disorder (age of onset) |
| MUTATION | Repeat or other mutation |
| SHORT | Number of CAG repeats normal allele |
| LONG | Number of CAG repeats longer allele |
| MUTATIONSPC | Specify mutation |
| TESTED | Tested relative |
| SCANCAT | Genotypes excluded from SCA unknown |
| ATAXIA | With SCA unknown ataxic yes / no |
| AOO2 | At SCA unknown AOO |
| SCACONTROL | Control group membership (family, community) |
| OLDSTUDY | Previous study |
| OLDID | Previous ID |
| HANDEDNESS | Handedness |
| SGAIT | Score SARA item Gait |
| SSTANCE | Score SARA item stance |
| SSITTING | Score SARA item sitting |
| SDISTURB | Score SARA item speech |
| SCHASERI | Subscore right SARA item finger chase |
| SCHASELE | Subscore left SARA item finger chase |
| SFINGERMEAN | Score SARA item finger chase |
| SFINGERRI | Subscore right SARA item nosefinger |
| SFINGERLE | Subscore left SARA item nosefinger |
| SFINGERNOSEMEAN | Score SARA item nosefinger |
| ALHANDRI | Subscore right SARA item diadochokinese |
| ALHANDLE | Subscore left SARA item diadochokinese |
| SALTHANDMEAN | Score SARA item diadochokinese |
| SHEELRI | Subscore right SARA item heelshin |
| SHEELLE | Subscore left SARA item heelshin |
| SHEELSHINMEAN | Score SARA item heelshin |
| SARASUM | SARA sum score |

**Supplementary Table 2. Variables used as a base set.**

# Supplementary Text

## Supplementary Text 1. Reasons outlining why OMOP was not suitable for dementia datasets

The OMOP Data Model, developed and maintained by **OHDSI**, is highly popular in bioinformatics since it provides a good standardization capability, while OHDSI also offers analysis tools that work with data in OMOP.

Since the beginning of the project, we have worked very closely with clinicians from the University Clinic of Bonn and the DZNE. During that time, we had insight into multiple different studies dealing with dementia and ataxia data. There was no common structure between those data sources and we quickly realized that in order to harmonize as much study data as possible we need a very flexible data model. Since we are working with data from one specific field (neurodegenerative diseases), we decided to go for a custom data model that does not come with the overhead of a data model that tries to capture all health observation data like OMOP does. Additionally there are some variables in the datasets we had access to that were not in OMOP and nor were there equivalent variables in OMOP. Such examples are:

- Neurological test questions like FAQFORM, FAQFINAN, FAQSHOP, etc. from the DZNE datasets
- Ataxia related variables like SGAIT, SSTANCE, SSITTING, etc
- ENTORHINAL_ICV, FUSIFORM_ICV, etc. from ADNI

## Supplementary Text 2. Implementation of the Data Steward Tool

Django is a high level web-framework written in Python that offers plugins for MongoDB as well as for providing RESTful APIs. The application holds the structure of the Clinical Data Model in terms of Python classes and is capable of communicating that to the underlying database. Uploaded data as well as mappings and the complete data logic is handled here. Moreover the application provides the APIs that can be queried via other systems. The Vue.js web application uses that API to yield a visual interface. Vue.js is a progressive Javascript framework to build single page applications like the Data Steward Tool. The communication between the Vue.js app and the Django backend is based on the RESTful API principle that uses JSON as notation for data exchange. The underlying MongoDB database stores the underlying data model with all its variables and mappings as well as normalized clinical data uploaded by the user. The deployment of the services is realized in a microservice architecture with Docker.

## Supplementary Text 3. Resources used for developing CDM

During the development of the Clinical Data Model we used different data sources in the form of clinical studies or patient registries. Most of the base variables are taken from, or influenced by, two major studies from the German Center for Neurodegenerative Diseases (DZNE). Those two studies are DESCRIBE DZNE - Clinical

Register Study of neurodegenerative Disorders[1] and DELCODE DZNE - Longitudinal Cognitive Impairment and Dementia study[2]. Additionally, we investigated international studies like Alzheimer's Disease Neuroimaging Initiative: ADNI[3] in order to be aligned with variables and terms used outside of the German research landscape. Other studies and (meta-) data resources that were used either during the development of the base variable set or in the later work to establish mappings were AddNeuroMed - the European collaboration for the discovery of novel biomarkers for Alzheimer's disease[4], terms from the Diagnostic and treatment center for memory disorders (DBGA) of UKB as well as multiple studies related to Ataxia we got access to from the DZNE: ESMI (European Spinocerebellar Ataxia Type 3/Machado-Joseph Disease Initiative)[5], SCA Registry (Registry for Spinocerebellar Ataxias (SCA))[6].

## Supplementary Text 4. Definition: Variable Mappings

A crucial part of the data model are variable mappings. The most common definition for that is a reference from one variable VAR1 to another one VAR2, where VAR1 and VAR2 are semantically equivalent. A single mapping in the CDM holds the information about the external variable VAR1 and the internal variable VAR2 plus the source of VAR1. All variable mappings can be viewed in tabular form: https://data-steward.bio.scai.fraunhofer.de/data-steward/table. With more and more mappings the system can read and understand data from many different data sources. In the future we expect the number of mappings (currently 276) to outgrow the number of internal variables (277) by far.

---

[1] https://www.dzne.de/en/research/studies/clinical-studies/describe-1-1/
[2] https://www.dzne.de/en/research/studies/clinical-studies/delcode/
[3] http://adni.loni.usc.edu/
[4] https://pubmed.ncbi.nlm.nih.gov/19906259/
[5] https://www.dzne.de/en/research/studies/clinical-studies/esmi/
[6] https://www.dzne.de/en/research/studies/clinical-studies/sca-registry/

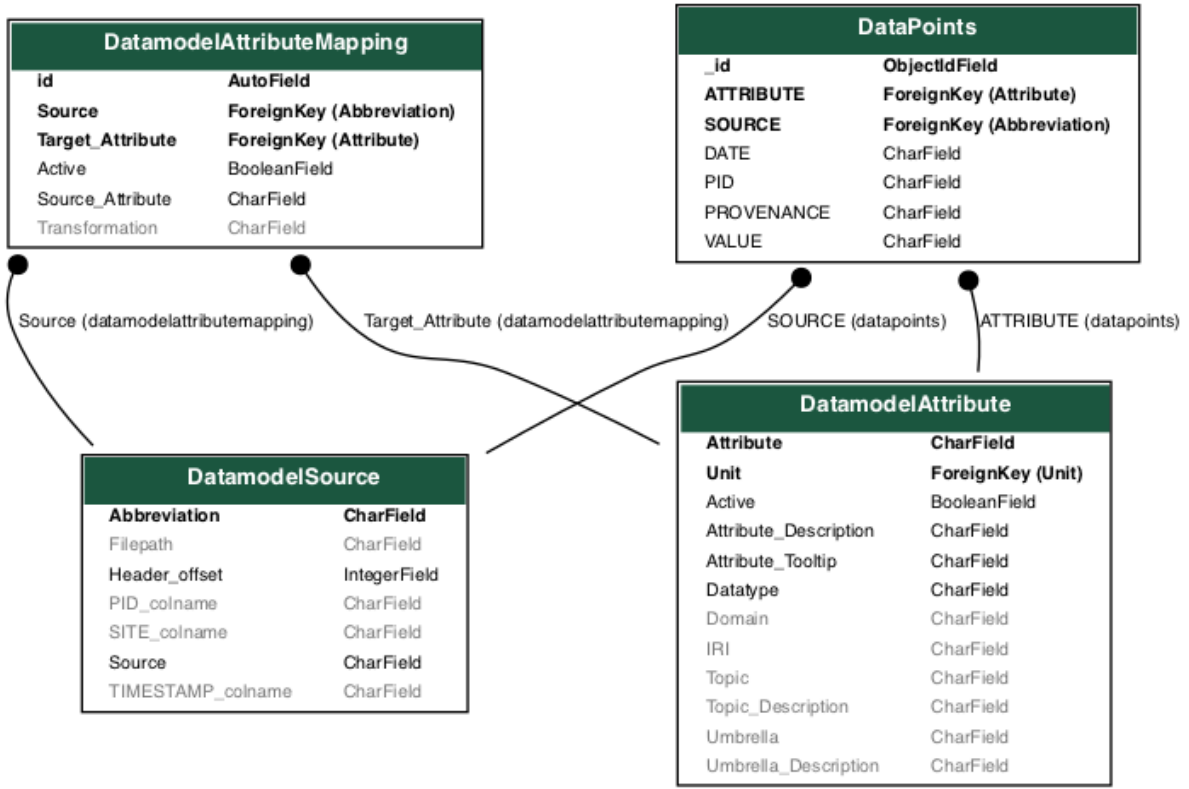

**Supplementary Figure 1. UML Diagram of CDM on Database level**

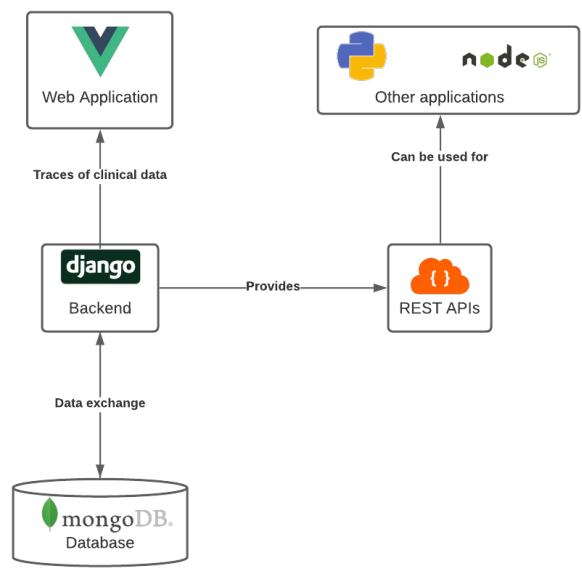

**Supplementary Figure 2. Interactions of the different Components of the DS.**

## Supplementary Text 5. Tutorial of the upload process

Clinical data is typically stored in 2D data tables, where in each row one measurement for one patient or subject is represented with the variable name and the measured value. The tool reads data in a csv-like format where each line represents one measurement. For example one line consists of the patient's id (Entity), the variable AGE (Attribute), and the patient's age (Value). A software transforming clinical data, that has multiple values per row, into the correct format is available. Such representation of clinical data is called EAV (**E**ntity **A**ttribute **V**alue) format. The Data Steward Tool can read and process those EAV files in a drag and drop section (**Figure 3**).

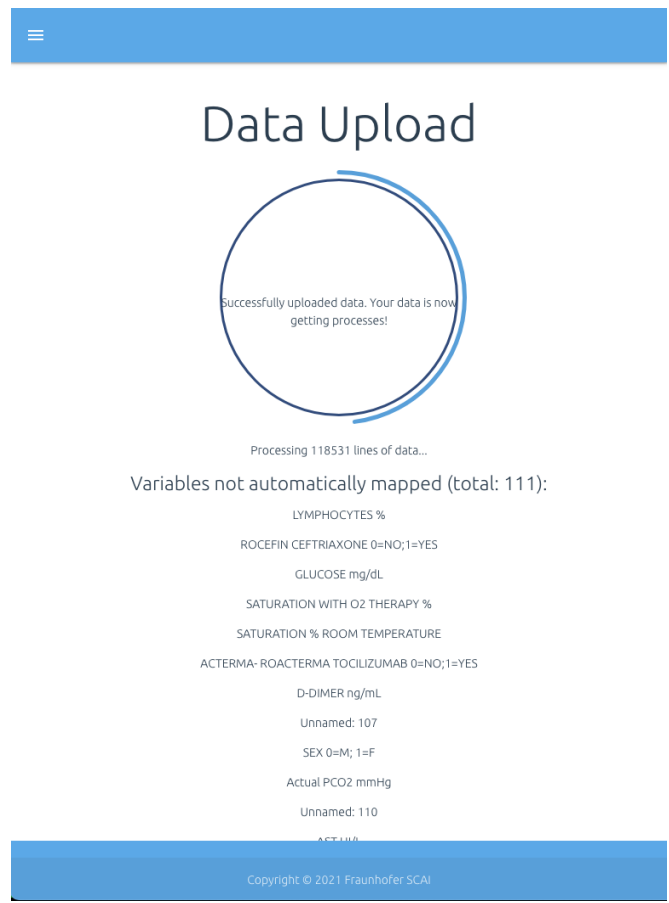

**Supplementary Figure 3**. **Data upload screenshot**

During the reading process, the user is updated with live feedback on how many lines in the file were found for how many distinct patients and what variables could not be found in the model. After finishing the process, the tool yields a summary of the results for the user to analyze. In the case that every variable was found in the underlying Data Model, everything is done and the data is successfully semantically integrated. If that's not the case the user can always map the variables onto CDM via a guided process. From the upload feedback, the user can go to the mapping assistant (**Figure 4**).

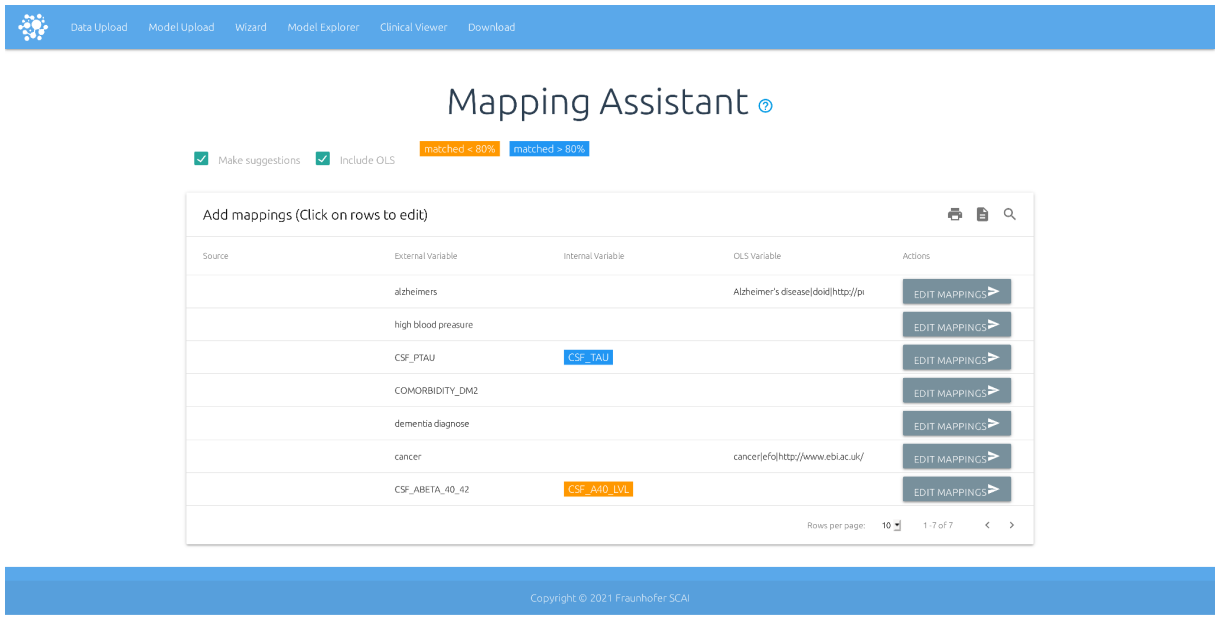

**Supplementary Figure 4. Mapping Assistant Screenshot.**

The assistant view is a table view where each line is an unmapped variable. The table comes with autocomplete and auto-suggestion features based on nearest neighbor search with edit distance as well as string comparison of variables and variable definition texts. Moreover, the user can activate the OLS (**O**ntology **L**ookup **S**ervice) autocomplete to conveniently integrate terms from major ontologies if no suitable variable is present in the current model. If a suitable variable was found the user has to select a source for the external variable (e.g. ADNI) and then the mapping can be submitted and the table gets reduced by the respective line. Thus, the user has a comfortable way to map all of their data onto the CDM.

## Supplementary Text 6. Aligning CDM with FHIR and OMOP

In bioinformatics, a very crucial part of research is exchanging health data between healthcare ecosystems or institutions. The FHIR (**F**ast **H**ealthcare **I**nteroperability **R**esources) standard of HL7 is a common solution to do that. Hence the Data Steward Tool can work as a FHIR server by providing multiple APIs (API Documentation[7]) that return either all patient data in FHIR format (observation resource defined by FHIR standard[8]) or single measurements for one patient. Moreover users can query the DS for certain patients in certain studies (patient resource[9]).

As mentioned above, the OMOP data model is an established solution for data standardization in bioinformatics and hence it is inevitable that the CDM is aligned with OMOP. Thus, we found mappings between every variable of the CDM and the OMOP standard vocabulary using OHDSI's Athena browser[10]. This embedding of the CDM onto OMOP is updated regularly if the CDM changes or after a certain time to keep the mappings up to date. Up to now, we were able to map 181 variables out of CDM to the OMOP standard vocabulary.

---

[7] https://data-steward.bio.scai.fraunhofer.de/data-steward/swagger
[8] https://www.hl7.org/fhir/observation.html
[9] https://www.hl7.org/fhir/patient.html
[10] https://athena.ohdsi.org/search-terms/start

## Supplementary Text 7. Importing other data models by the example of i2b2

As mentioned in the paper there are some major data models out there that aim to capture general biomedical data like OMOP or domain specific data like GA4GH. The i2b2 data model is capable of describing general clinical data. In order to contribute to a research landscape where data interoperability is crucial it is important that the data steward tool is able to function with other data models in its backend. We have created a python package available on GitHub[11] that contains an example of how to import all i2b2 variables into the DST.

## Supplementary Text 8. Duplicating virtual patients

We manually generated 10 virtual patients from ADNI data using a mixed technique consisting of statistical bootstrapping and adding normal distributed noise to the observation values. The resulting data, and with that the whole method, was evaluated and tested using the currently known data ranges from the original 'true' ADNI data from the study. By that, we could generate artificial data that we could use for public instances without any privacy concerns while also demonstrating realistic data values. Results and implementations are available on GitHub[12].

---

[11] https://github.com/phwegner/DST-API
[12] https://github.com/phwegner/AAD_N_DR